\documentclass[preprint,preprintnumbers,amsmath,amssymb,nofootinbib,aps]{revtex4}
\pdfoutput=1
\usepackage{epsfig}

\usepackage{graphicx}
\usepackage{dcolumn}
\usepackage{bm}
\usepackage{amsmath}
\usepackage{latexsym}
\usepackage{color}
\usepackage{hyperref}
\usepackage{mathrsfs}

\newcommand{\be}{\begin{equation}}
\newcommand{\ee}{\end{equation}}

\begin{document}


\title{An unified cosmological evolution driven by a mass dimension one fermionic field}

\author{S. H. Pereira$^{1}$} \email{s.pereira@unesp.br}
\author{M. E. S. Alves$^{2,1}$} \email{marcio.alves@unesp.br}
\author{T. M. Guimar\~aes$^{3,1}$}\email{thiago.mogui@gmail.com}

\affiliation{$^1$Universidade Estadual Paulista (UNESP), Faculdade de Engenharia de Guaratinguet\'a, Departamento de F\'isica e Qu\'imica, Guaratinguet\'a, SP, 12516-410, Brazil
\\$^2$Universidade Estadual Paulista (UNESP), Instituto de Ci\^encia e Tecnologia, S\~ao Jos\'e dos Campos, SP, 12247-004, Brazil
\\$^3$Instituto Federal do Paran\'a, Campus Ivaipor\~a,
Ivaipor\~a, PR, 86870-000, Brazil
}

\begin{abstract}
An unified cosmological model for an Universe filled with a mass dimension one (MDO) fermionic field plus the standard matter fields is considered. After a primordial quantum fluctuation the field slowly rolls down to the bottom of a symmetry breaking potential, driving the Universe to an inflationary regime that increases the scale factor for about 71 e-folds. After the end of inflation, the field starts to oscillate and can transfer its energy to the standard model particles through a reheating mechanism. Such a process is briefly discussed in terms of the admissible couplings of the MDO field with the electromagnetic and Higgs fields. We show that even if the field loses all its kinetic energy during reheating, it can evolve as dark matter due a gravitational coupling (of spinorial origin) with baryonic matter. Since the field acquires a constant value at the bottom of the potential, a non-null, although tiny, mass term acts as a dark energy component nowadays. Therefore, we conclude that MDO fermionic field is a good candidate to drive the whole evolution of the Universe, in such a way that the inflationary field, dark matter and dark energy are described by different manifestations of a single field.
\end{abstract}

\maketitle
\section{Introduction}
\label{sec:introd}

A new class of mass dimension one fermionic fields has been proposed by Ahluwalia and Grumiller \cite{AHL1,AHL2,ahl2011a}, which are constructed by means of charge conjugation spinors. In its first formulation the spin sums of the quantized fields showed to be Lorentz violating due to the intrinsic definition of their dual fields, which rose several doubts on the applicability of this new field. Such an old construction were named Elko\footnote{{\it Eigenspinoren des ladungskonjugationsoperators}, or Self-spinor of charge conjugation} fields.  However, recently \cite{AHL4,AHL3} a subtle deformation in the dual structure solved the problems of Lorentz violation, putting the theory in solid bases from the point of view of the quantum field theory, and the new fields are called just mass dimension one (MDO) fermionic fields. In this work we are interested just in the classical analogue of MDO fermionic fields, thus we can use Elko or MDO fields without loss of generality.

{The fermionic fields constructed from charge conjugation spinors are natural candidates to describe dark matter particles in the universe, since they are neutral and have canonical mass  dimension one, which make them to couple very weakly to other particles of the standard model. Indeed, the only admissible couplings are with scalars and Higgs fields \cite{alves1,dias_lee,AlvesDias2014} and with electromagnetic stress tensor \cite{alves2}, in addition to quartic self-couplings. Different from Dirac fermion, which has mass dimension ${3\over 2}$, is parity conjugation invariant and admits several couplings with standard model particles, the MDO field born with an intrinsic dark character. Consequently, detection and constraints in its physical properties is difficult, once it does not couple directly with standard model particles.}

Several cosmological applications of this new field have been done, first in torsion free frameworks \cite{FABBRI,BOE4,BOE6,GREDAT,BASAK,basak2015,sadja,saj,js,asf} and more recently considering its coupling to torsion \cite{kouwn,sajf,st,sra} in an Einstein-Cartan framework, which turns the system of equations richer than the previous case, especially for inflationary applications. In particular, in Refs. \cite{st,sra} the numerical results describing inflation and dark matter evolution were presented together with the energy density evolution, in good agreement with the required inflationary number of e-folds and also with the expected energy density for the dark matter component. Notwithstanding, we refer the reader to other important recent results regarding the MDO fermionic field in the context of the quantum field theory \cite{dias1,elkosigma,elkocasimir,elkoeffective}, and also in thermal field theory \cite{sarich}. {The study of scalar and tensor perturbations for Elko field has been done in \cite{GREDAT,BASAK} in the torsion free case and the  study of first
order vector perturbations has been done in \cite{basak2015}.}
 
In the present article we consider the fermionic MDO field as a candidate to drive the complete evolution of the Universe. A symmetry breaking potential is responsible for the inflationary phase, while a reheating like mechanism drives the field to the radiation phase. The effective mass of the field plays an essential role in the energy transfer process. After the field rolls down to the true vacuum of the potential, the system follows a dark matter evolution due to a natural (gravitational) coupling between the MDO field and the baryonic energy density. Finally, a dark energy accelerated phase can be obtained from the constant quadratic mass term, which acts exactly like a cosmological constant term. All different phases are connected and the free parameters can be constrained by observational data and by some reasonably physical assumptions. The cosmic coincidence problem can also be naturally understood in such a scenario.

It is worth to stress that the search for an unified cosmological model is an old task and most of the models are based on a single scalar field \cite{Liddle2006,Liddle2008,Bastero2016}. The advantage to use scalar fields is that the inflationary phase and the generation of primordial perturbations are in good agreement with observations \cite{bookliddle,linde1,kolb}. After reheating, if the decay of the scalar field is incomplete, it may act as dark matter, while its zero point energy acts as dark energy. Although this is a standard model framework, scalar fields does not furnish any physical interpretation for inflation and also no scalar field has been observed yet, except the Higgs boson. 

On the other hand, the general idea of considering the usual Dirac fermions to explain inflation, dark matter and dark energy have already been considered in the literature (see, e.g., \cite{saha1997,saha2004,saha2006,grams2014}). In this spirit, the aim of the present article is to show that the MDO fermionic field is a more reasonable field to drive inflation and the subsequent stages of evolution of the Universe in an unified fashion. Being neutral and not interacting with other particles of the standard model, it is a good candidate to describe dark matter without the need of supposing an incomplete decay during reheating, an it happens in the case of scalar dark matter particles. 

Furthermore, as pointed out by Pereira et al. \cite{st,sra}, the ``MDO inflation'' can be interpreted in light of the Pauli exclusion principle. When the fermionic field rolls down to the minimum energy state of the potential, the Pauli exclusion principle starts to act, not allowing all particles to occupy the lowest energy state. If the potential is stronger than the degeneracy pressure, the whole system responds with an abrupt expansion in order to accommodate all particles in the lowest energy state, since the spacing between energy levels in a bound system is inversely proportional to the size of the system. This effect allows all the particles to accommodate very close to the lower energy state after inflation.

Moreover, since the MDO particles satisfy a Klein-Gordon like equation, the set of equations describing the evolution of the Universe is very similar to the scalar field case. Thus, several features of the scalar field cosmology are recovered, but for a fermionic field instead.

The paper is organized as follows. Section \ref{sec 2} presents the basic Friedmann-Lema\^itre-Robertson-Walker (FLRW) equations already derived in \cite{sajf,st,sra} in the presence of torsion terms. The dark matter and the dark energy behavior of the MDO field are presented in the Section \ref{sec 3}, where some parameters are constrained with observational data. In the Section \ref{sec 4} the MDO inflation is studied by means of a numerical solution of the evolution of the field subject to a symmetry breaking potential. Two possible ways of implement the reheating phase in the MDO cosmological scenario is briefly discussed in the Section \ref{sec 5}, and in the Section \ref{sec 6} we finish with our conclusions.

\section{Dynamic equations for the MDO fermionic field}\label{sec 2}

The action for the model is \cite{kouwn,sajf}
\begin{equation}\label{actionE}
S = \int d^4 x \sqrt{-g} \left[ -\frac{\tilde{R}}{2\kappa^2}+{1\over 2}g^{\mu\nu}\tilde{\nabla}_\mu \stackrel{\neg}{\Lambda}\tilde{\nabla}_\nu \Lambda -V(\stackrel{\neg}{\Lambda}\Lambda) \right] + S_m,
\end{equation}
where $\kappa^2\equiv8\pi G=8\pi/m_{pl}^2$ with $c= \hbar = 1$ and $S_m$ is the usual action for other matter fields, as baryonic matter or radiation. The tilde represents torsion terms into the Ricci scalar $\tilde{R}$ and covariant derivatives, namely, $\tilde{\nabla}_\mu \Lambda\equiv \partial_\mu\Lambda - \tilde{\Gamma}_\mu \Lambda$ and $\tilde{\nabla}_\mu \stackrel{\neg}{\Lambda} \equiv \partial_\mu\stackrel{\neg}{\Lambda} + \stackrel{\neg}{\Lambda}\tilde{\Gamma}_\mu$,
where $\tilde{\Gamma}_\mu$ are the spin connections. In the Einstein-Cartan framework the contorsion terms generalizes the affine connection $\tilde{\Gamma}^\rho_{\mu\nu} = \Gamma^\rho_{\mu\nu}+K^\rho_{\;\;\mu\nu}$ and are given by
\begin{equation}
K^\rho_{\;\;\mu\nu}=-{1\over 2}(T^\rho_{\;\;\mu\nu}+T_{\mu\nu}^{\;\;\;\;\rho}+T_{\nu\mu}^{\;\;\;\;\rho} )\,,
\end{equation}
where the only non-vanishing torsion terms obeying cosmological principle are \cite{tsam}
\begin{equation}
T_{ii0} = - T_{i0i} = a(t)^2h(t),~~~~ i=1,\,2,\,3
\end{equation}
\begin{equation}
T_{ijk}=2a(t)^3f(t)\varepsilon_{ijk},
\end{equation}
where the torsion functions $h(t)$ and $f(t)$ must be determined. 

For the MDO fermionic field we have used just one of the four different fields represented by $\Lambda$, satisfying $\Lambda=\phi(t) \lambda$, where $\lambda$ represents the fermionic field in a Minkowski space-time which is normalized as $\stackrel{\neg}{\lambda}\lambda=1$, and $\phi(t)$ carries its time evolution in a Friedmann-Lema\^itre-Robertson-Walker (FLRW) background \cite{BOE4,BOE6,GREDAT,
BASAK,sadja,kouwn,sajf,saj,js,asf,st,sra}. 

Considering the flat FLRW metric $ds^2=N(t)^2 dt^2-a(t)^2[dx^2 + dy^2 + dz^2]$, where $N(t)$ is the lapse function, the two Friedmann equations, the dynamic field equation for $\phi(t)$ and the torsion functions $h(t)$ and $f(t)$ can be obtained by taking the variation of the Lagrangian of the model with respect to
$N(t)$, $a(t)$, $\phi(t)$, $h(t)$ and $f(t)$ respectively (see \cite{kouwn,sajf,st,sra} for further details). Thus we obtain (setting $N\to 1$ at the end)
\begin{equation}
H^2={\kappa^2\over 3}\bigg(1+{\kappa^2\phi^2\over 8} \bigg)\bigg[{\dot{\phi}^2\over 2}+V(\phi) + \rho_m\bigg]\,,\label{H2}
\end{equation}
\begin{equation}
\dot{H}=-{\kappa^2\over 2}\bigg(1+{\kappa^2\phi^2\over 8} \bigg)\bigg[\dot{\phi}^2-{1\over 2}{H\phi\dot{\phi} \over (1+\kappa^2\phi^2/8)^2} +\rho_m + p_m\bigg]\,,\label{Hdot}
\end{equation}
\begin{equation}
\ddot{\phi}+3H\dot{\phi}+{dV(\phi)\over d\phi}-{3\over 4}{H^2\phi\over (1+\kappa^2\phi^2/8)^2}=0\,,\label{phiElko}
\end{equation}
\begin{equation}
h(t)=-{1\over 8}{\kappa^2\phi^2\over (1+\kappa^2\phi^2/8)}H(t)\;, \quad f(t) = 0\,,\label{torsion}\\
\end{equation}
where $H=\dot{a}/a$, as usual, and $\rho_m$ and $p_m$ are the energy density and the pressure of other matter components, which satisfies  a conservation equation for a perfect fluid
\begin{equation}
\dot{\rho}_m + 3H(\rho_m + p_m)=0.
\end{equation}

On the other hand, the energy density and the pressure of the MDO field is given by
\begin{equation}
\rho_\phi={\dot{\phi}^2\over 2}+V(\phi)+{3\over 8}{H^2\phi^2\over (1+\kappa^2\phi^2/8)}\,,\label{rhoElko}
\end{equation}
\begin{eqnarray}
p_\phi&=&{\dot{\phi}^2\over 2}-V(\phi)-{3\over 8}{H^2\phi^2\over (1+\kappa^2\phi^2/8)}- {1\over 4 }{\dot{H}\phi^2\over (1+\kappa^2\phi^2/8)}-{1\over 2}{H\phi\dot{\phi}\over (1+\kappa^2\phi^2/8)^2}\,.\label{pphiElko}
\end{eqnarray}

The set of Eqs. (\ref{H2})-(\ref{pphiElko}) looks-like a generalization of the standard scalar field model,  but it is important to keep in mind that here $\phi(t)$ is just the temporal part of the fermionic field $\Lambda$. Now, by substituting the Eqs. (\ref{H2}) and (\ref{Hdot}) in the Eqs. (\ref{rhoElko}) and (\ref{pphiElko}), it is possible to write the above expressions of $\rho_\phi$ and $p_\phi$ in a different useful form
\begin{equation}\label{Elko density new}
\rho_\phi = \frac{\kappa^2 \phi^2}{8}\rho_m + \left(1+\frac{\kappa^2\phi^2}{8}\right)\left[\frac{\dot{\phi}^2}{2} + V(\phi)\right],
\end{equation}
\begin{equation}\label{Elko pressure new}
p_\phi = \frac{\kappa^2 \phi^2}{8} p_m + \left(1+\frac{\kappa^2\phi^2}{8}\right)\left[\frac{\dot{\phi}^2}{2} - V(\phi)\right]  - \frac{1}{2}\frac{H\phi \dot{\phi}}{(1+ \kappa^2 \phi^2/8)},
\end{equation}
which can be straightforwardly combined to show that
\begin{equation}\label{phiElko2}
\dot{\rho}_\phi + 3H(\rho_\phi + p_\phi)=0.
\end{equation}

Notice also the presence of a coupling between the MDO field with the energy density and pressure of the standard matter into the first terms of Eqs.  (\ref{Elko density new}) and (\ref{Elko pressure new}). Such terms are a manifestation of the coupling between the spin components of the MDO field with gravity, such that if one takes the Minkowski spacetime ($H=0$) then $\rho_\phi$ and $p_\phi$ for the scalar field are exactly recovered, as can be directly checked by means of the Eqs. (\ref{rhoElko}) and (\ref{pphiElko}).

From the above formulation, we see that the present model is equivalent in form to a model with zero torsion containing only perfect fluid components, such that one of them has the energy density and pressure defined by the Eqs. (\ref{Elko density new}) and (\ref{Elko pressure new}). Moreover, the Eq. (\ref{phiElko2}) is obviously equivalent to the Eq. (\ref{phiElko}), and the Eqs (\ref{H2}) and (\ref{Hdot}) can now be written as
\begin{equation}
H^2={\kappa^2\over 3}(\rho_\phi + \rho_m)\,,\label{H22}
\end{equation}
\begin{equation}
\dot{H}=-{\kappa^2\over 2}(\rho_\phi + p_{\phi} + \rho_m+p_m).\label{Hdot2}
\end{equation}

Regarding the potential $V(\phi)$, two forms were previously considered  in the literature. The first one is a symmetry breaking potential \cite{st}
\begin{equation}
V_1(\phi)=A^4\Bigg(1-\frac{\phi^2}{\phi_c^2}\Bigg)^2 \,,\label{V1}
\end{equation}
where  $A$ and $\phi_c$ are positive constants. It was showed that as the field rolls down to the true vacuum  of the potential at $\phi = \phi_c$ the inflation occurs with the correct number of e-folds, depending on the initial value of $\phi$ and also on the constants $A$ and $\phi_c$. After inflation, a dark matter evolution follows naturally, leading to correct energy densities for different phases.

On the other hand, in the Refs. \cite{sajf,sra} the potential was chosen to be of the form
\begin{equation}
V_2(\phi)= \frac{1}{2}m^2 \phi^2+ {\alpha\over 4}\phi^4\,,\label{V2}
\end{equation}
where $m$ is the physical mass of the field and $\alpha$ is a dimensionless coupling constant. Considering such a potential, the inflationary and dark matter evolution were also obtained \cite{sra}, with the correct numerical estimates to the energy density at different epochs. Furthermore, a scenario where $\phi(t)$ is a slowly varying function at late times was interpreted as a time varying cosmological term proportional to $H^2$ \cite{sajf}.

Despite the successful to describe some individual phases of evolution with correct numerical estimates, the radiation phase and a smooth transition to late cosmic acceleration were not completely addressed by the model. In order to describe all the phases of evolution of the Universe in a consistent unified fashion, here we will consider a potential of the form
\begin{equation}
V(\phi)=V_1(\phi)+V_2(\phi)\,,\label{Vtotal}
\end{equation}
with $V_1$ and $V_2$ given by (\ref{V1}) and (\ref{V2}), and also the presence of the standard matter fields. As it will be seen in the following sections, all the stages of evolution of the Universe can be recovered in a natural way.

{In order to better understand the role of the above potential in the dynamic of the field let us write it in the expanded form as
\begin{equation}
V(\phi)= \frac{1}{2}m_{eff}^2 \phi^2+ {\bar{\alpha}\over 4}\phi^4 + C\,,\label{Vexp}
\end{equation}
where $m_{eff}^2=m^2-4A^4/\phi_c^2$ represents an effective mass of the field, $\bar{\alpha}=\alpha+4A^4/\phi_c^4$ is an effective self coupling constant and $C=A^4$ a constant. 
The potential is attractive whatever the $\phi$ value if $\bar{\alpha}>0$, which leads to $\alpha > -4A^4/\phi_c^4$, and
the minimum occurs at 
\begin{equation}
\phi_{min}=\phi_c \sqrt{\frac{1-m^2\phi_c^2/4A^4}{1+\alpha \phi_c^4/4A^4}}\label{phimin}
\end{equation}
For $m^2<<4A^4/\phi_c^2$ and $\alpha<<4A^4/\phi_c^4$ the minimum occurs at $\phi_{min}\approx \phi_c$, which will be the case when we fix the parameters with observational constraints.}

\section{Dark matter and dark energy evolution}\label{sec 3}

From now on we will consider that the only additional standard matter present is of pressureless baryonic type, namely $\rho_m = \rho_b$ and $p_m=p_b=0$. Let us start with the late time evolution of the Universe in order to constrain the parameters $\phi_c$, $m$ and $\alpha$ with observational data.  The value of $\phi_c$ is of particular interest in the inflationary epoch for which we give a detailed description in the next section. For now, it is enough to suppose that, at early times, the field $\phi$ is initially at rest around the false vacuum of $V_1$ and that after a quantum fluctuation, it rolls down to the bottom of the potential acquiring, after a period long enough, the constant value $\phi_c$. Hence, when the kinetic energy of the MDO field is finally negligible, with the field satisfying $\dot{\phi}\simeq 0$, we have $V_2>> V_1$. Therefore, at late times when only the baryonic matter and the MDO field are relevant for the cosmic dynamics, we can use the Eqs. (\ref{Elko density new}) and (\ref{Elko pressure new}) to obtain
\begin{equation}\label{Elko density baryon}
\rho_\phi = \frac{\kappa^2 \phi_c^2}{8}\rho_b + \left(1 + \frac{\kappa^2 \phi_c^2}{8} \right)V_2(\phi_c),
\end{equation}
and
\begin{equation}
p_\phi = - \left(1 + \frac{\kappa^2 \phi_c^2}{8} \right)V_2(\phi_c),
\end{equation}
where $\rho_b$ is the baryonic energy density and the potential has the fixed value $V_2(\phi_c) = \frac{1}{2} m^2\phi_c^2 + \frac{1}{4} \alpha\phi_c^4$. From the above expressions, it is clear that at late times the energy density of the MDO field is an addition of two distinct contributions. The first one behaves as a pressureless fluid following the evolution of the baryonic matter, and the second one is an effective cosmological constant given by
\begin{equation}
\Lambda_{\rm eff} \equiv \kappa^2\left(1+ \frac{\kappa^2 \phi_c^2}{8}\right)V_2(\phi_c).\label{Leff}
\end{equation}

Inserting $\rho_\phi$ and $\rho_b$ into the Friedmann equation (\ref{H22}), the Hubble function can be obtained as follows
\begin{equation}\label{hubble}
\frac{H(a)^2}{H^2_0} = \frac{\Omega_b}{a^3} + \frac{\Omega_{{\rm DM,}\phi}}{a^3} + \Omega_{\Lambda,\phi},
\end{equation}
where\footnote{The subscript 0 stands for the present day quantities and $\rho_{{\rm crit},0} = 3H_0^2/\kappa^2$.} $\Omega_b = \rho_{b,0}/\rho_{{\rm crit},0}$ and now it is clear that the second term in the right-hand-side of the above equation can be interpreted as a dark matter component with the present density parameter defined as
\begin{equation}\label{relation dark barion}
{\Omega_{{\rm DM,}\phi}} \equiv \frac{\kappa^2\phi_c^2}{8} \Omega_b,
\end{equation}
which comes from the gravitational coupling of the MDO field with standard baryonic matter. The last term act as a cosmological constant term, with $\Omega_{\Lambda,\phi}\equiv\Lambda_{\rm eff}/\kappa^2 \rho_{{\rm crit},0}$. Notice that Eq. (\ref{hubble}) has exactly the same form as the $\Lambda$CDM model, with baryonic and dark matter components following the same evolution as $a^{-3}$.

At the end of the inflationary and reheating stages, gravity is the only non-negligible interaction of the fermionic MDO field (besides its self-interaction) which can cluster, enabling the posterior formation of structures in the Universe. The above definition of the dark matter density parameter of MDO shows a remarkable feature of this field, namely, its natural coupling with other matter fields which comes directly from the structure of the equations of motion in curved spaces. It is worth to stress that we have not included any direct coupling in the action (\ref{actionE}). Due to its gravitational origin, such a coupling disappears in the Minkowski space.

The usual standard scalar field can also behaves as dust matter, and unified inflation-dark matter scenarios have been explored in the literature (see, e.g., \cite{Liddle2006,Bastero2016} and references therein). But in this case, the zero pressure is obtained by a time average of the field while it is coherently oscillating after the end of the slow-roll regime. Nevertheless, a dark matter contribution can only survive if the decay of the inflaton is incomplete. The MDO field also oscillates at the end of inflation as we will see in the Section \ref{sec 5}, however it can evolve as dark matter at late times even if its kinetic energy decays completely during reheating. This is explained by the presence of the term evolving as $a^{-3}$ in the Eq. (\ref{hubble}) coming from the MDO energy density. Therefore, the scenario described in this Section would not be possible if the field goes to zero after inflation as, e.g., in the chaotic MDO inflation studied in the Ref. \cite{sra}, since the coupling of the MDO field with the baryonic energy density is cancelled [see Eq. (\ref{Elko density baryon})]. The dark matter behaviour in this case is obtained by averaging the coherent oscillations of the MDO field in the present epoch leading to a zero pressure, in a similar manner to what happens to the scalar field inflationary model. 

Notwithstanding, assuming that the symmetry breaking potential $V_1(\phi)$ was dominant in the inflationary regime, the MDO potential acquires its true vacuum value as the amplitude of oscillations becomes virtually zero. Although small, such zero point energy leads to a non-negligible contribution today, which results in a non-null cosmological constant like parameter $\Omega_{\Lambda,\phi}$.

Finally, in order to estimate the values for the parameters of the theory, let us now consider, for instance, the last results from the Planck collaboration on the cosmological parameters \cite{Aghanim2018}. Considering a flat spatial section and $H_0 = 67.4~{\rm km}~{\rm s}^{-1}{\rm Mpc}^{-1}$ we have $\Omega_b \simeq 0.05$, $\Omega_{{\rm DM},\phi} \simeq 0.27$ and $\Omega_{\Lambda,\phi} = 1 - \Omega_b - \Omega_{{\rm DM},\phi} \simeq 0.68$. With these values in the relation (\ref{relation dark barion}) we obtain $\phi_c \simeq 1.3~m_{pl}$, and the value of the potential at its minimum can be obtained from (\ref{Leff}) and (\ref{relation dark barion}) in terms of the cosmological parameters as
\begin{equation}
V_2(\phi_c) = \left(\frac{\Omega_{\Lambda,\phi}}{1+\Omega_{{\rm DM},\phi}/\Omega_b }\right)\rho_{{\rm crit},0} \simeq 4.1 \times 10^{-48}~{\rm GeV}^4.\label{V2C}
\end{equation}
{From (\ref{Leff}) we obtain directly the value of the effective cosmological constant, namely $\Lambda_{eff} = 4.37\times 10^{-66}$eV$^2$, almost exactly the value obtained from Planck collaboration\footnote{From \cite{Aghanim2018}, $\Lambda = (4.24\pm 0.11)\times 10^{-66}$eV$^2$.} \cite{Aghanim2018}.}

{From (\ref{V2C}), (\ref{V2}) and the values of $\alpha$ and $\bar{\alpha}$ from (\ref{Vexp}) and (\ref{phimin}) we can also constraint the value for the physical mass of the field. With $
A\sim 10^{14}$ GeV (see next section) and $\phi_c \simeq 1.3~m_{pl}$, the $\alpha$ parameter must be in the interval $-10^{-18} \lesssim \alpha \lesssim 10^{-125}$. Thus, the range for the mass is $0 < m \lesssim 2.4 \times 10^{10}$ GeV. It is important to remark that, contrary to scalar field dark matter models \cite{Liddle2006}, here the contribution to dark matter comes just from the value of the incomplete field decay, namely $\phi_c$, through (\ref{relation dark barion}), and not from the mass $m$ of the field. The mass and $\alpha$ parameter are related to the cosmological constant like term, $\Omega_{\Lambda,\phi}$, which represents the dark energy sector. This opens the possibility to mass be in a very large range, which must be constrained by other methods, as the growth of primordial perturbations.}


\section{MDO inflation with a symmetry breaking potential}\label{sec 4}

{ Now, we can ask ourselves if the energy density of the MDO fermionic field can be dominant at the very early stages of evolution of the Universe. If we remember that one of the admissible couplings of the MDO field $\Lambda$ is with the Higgs field $\Phi$, we may give an affirmative answer as follows. A scenario analogous to the hybrid inflationary model \cite{Linde1993} can be pictured out for the present case. An interaction between $\Lambda$ and $\Phi$ makes the latter goes to zero rapidly, while the $\Lambda$ energy density could remain large for a much longer time since $\Lambda$ does not interact appreciably with other fields. In the present article, nevertheless, we do not consider the details of such a coupling, but we just consider the stage of inflation at large energy density for the MDO field while $\Phi = 0$, as it has already been discussed in the literature by one of us and other authors \cite{st,sra}. The coupling with the Higgs field could be particularly important in the last stage of inflation, and we intend to take this into account in forthcoming investigations.} 

Therefore, let us analyze the inflationary epoch by assuming that the energy density of the Universe is initially dominated by the MDO field. At the high energy regime of inflation the only relevant potential is the symmetry breaking potential $V_1(\phi)$. At the time that inflation begins, $t_i$, the initial value of the field $\phi_i \ll \phi_c$, and the constant $A$ can be chosen as $A \simeq (3H_i^2 m_{pl}^2/8\pi)^{1/4} \sim 5.3 \times 10^{14}$ GeV, such that the Hubble parameter assumes the following value $H(t_i) \equiv H_i \sim 10^{35}~{\rm s}^{-1}$ \cite{kolb} at the GUT scale. Conversely, it is remarkable that the constant $\phi_c$ is not fixed by inflation, but by the ratio between the amount of dark matter and baryonic matter in the Universe as it was shown in the last section [Eq. (\ref{relation dark barion})].

\begin{figure} 
\centering
\includegraphics[scale=1.2]{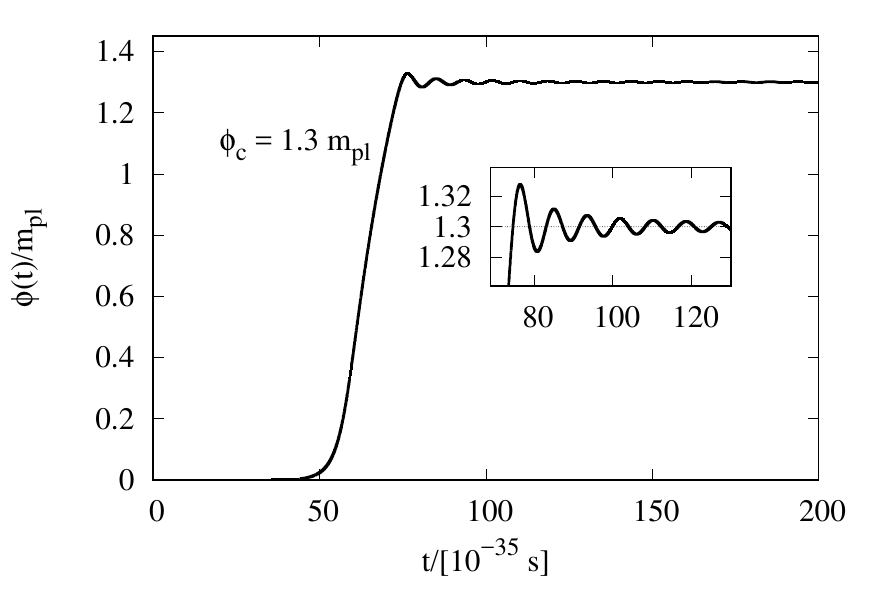}
\caption{Evolution of $\phi(t)$ during and after inflation. The inset graphic shows in detail the first oscillations of the field. }\label{fig elko oscil}
\end{figure}

The MDO field is initially in a nearly equilibrium point of the false vacuum of $V_1(\phi)$. Since the field has mass dimension one (or energy dimension), we can use a quantum uncertainty relation in order to establish its initial value. A quantum fluctuation $\Delta \phi$ in a time interval $\Delta t$ must satisfy $\Delta \phi  \Delta t  \geq 1$. Since inflation starts at $t_i \sim 10^{-35}$ s, we can choose $\Delta t$ of this order to arrive at $\Delta \phi \sim 1/t_i \sim 5.4 \times 10^{-9}~ m_{pl}$, which will be assumed as the initial condition for the field in what follows. 

On the other hand, the initial value of the first derivative of the field $\dot{\phi}(t_i) \equiv \dot{\phi_i}$ can be obtained by taking the equations of motion (\ref{phiElko}) at the time $t_i$. Let us consider that the condition $\ddot{\phi} \ll H\dot{\phi}$ is satisfied when inflation starts, then we are lead to
\begin{equation}
\dot{\phi}_i \simeq \frac{H_i\phi_i}{4} \left[1 + \frac{2}{\pi (\phi_c/m_{pl})^2} \right].
\end{equation} 
If one takes $\phi_c = 1.3~m_{pl}$, for instance,  the initial value is $\dot{\phi}_i = 1.86 \times 10^{26}~m_{pl}~{\rm s}^{-1}$.

Having settled the initial conditions, one can numerically solve the system of differential equations (\ref{H2})-(\ref{torsion}) in order to study the evolution of the dynamical parameters of the system. In this Section we consider just the MDO field as a source of energy, not including baryonic matter or radiation, which are assumed to be not dominant in that phase. 

In the Fig. \ref{fig elko oscil}, the evolution of the field $\phi$ is shown as a function of the cosmic time $t$. From this figure it is clear that the field evolves from $\phi_i$ to $\phi_c$ while rolling down to the bottom of the potential, with small oscillations at the end of its evolution, which will be discussed in the next Section. 

In order to study the duration and the kinematics of the inflation, it is useful to analyze the evolution of the slow roll parameters governing the inflationary epoch. They are defined in the usual way as
\begin{equation}
\epsilon \equiv \frac{|\dot{H}|}{H^2},~~~{\rm and}~~~\eta \equiv \frac{\ddot{\phi}}{H \dot{\phi}}~~.
\end{equation}

Moreover, it is convenient to analyze the evolution of the relevant quantities as functions of the number of e-folds achieved at the time $t$, defined as follows
\begin{equation}
N = \int_{t_i}^t H(t^\prime) dt^\prime.
\end{equation}

\begin{figure} 
\centering
\includegraphics[width=0.49\textwidth]{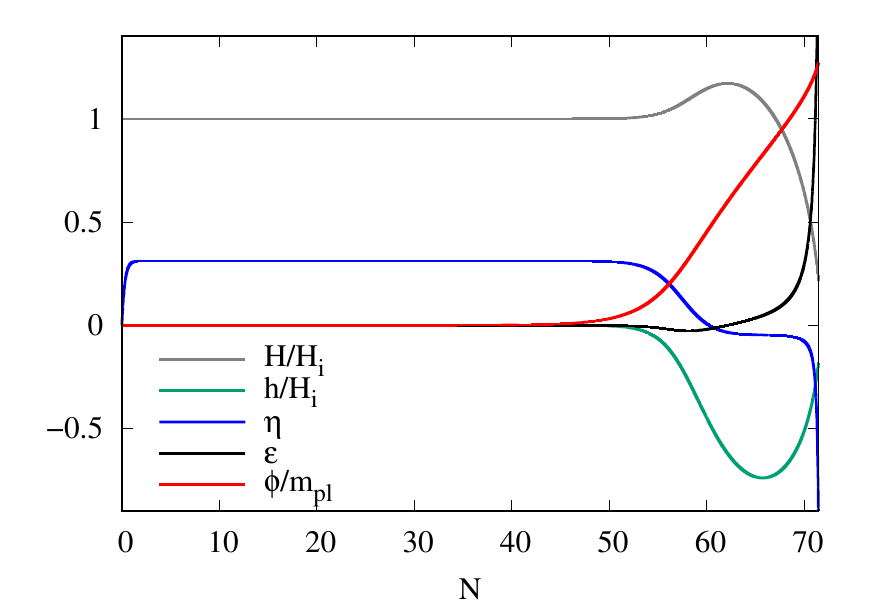}
\includegraphics[width=0.49\textwidth]{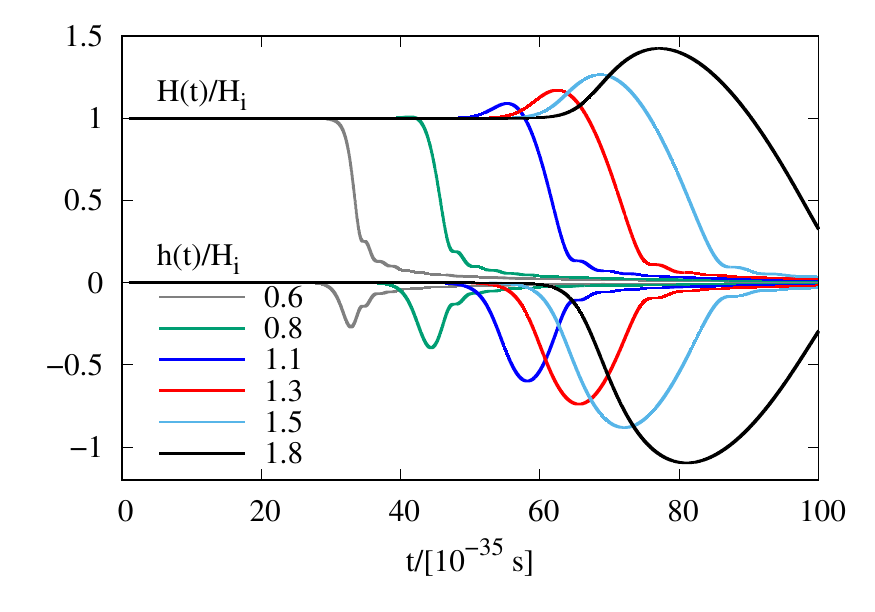}
\caption{(a) Evolution of the Hubble function $H$, the torsion function $h$, the slow-roll parameters $\epsilon$ and $\eta$, and of the field $\phi$ as functions of the number of e-folds during inflation. When the inflation ends $\epsilon = 1$ and we have $N \simeq 71$. All the quantities were obtained using $\phi_c = 1.3~m_{pl}$. (b) Evolution of $H$ as a function of the cosmic time $t$. The size of the bump at the end of inflation depends strongly on the value of $\phi_c$ (specified in the legend in units of Planck mass). Such a bump is related with the torsion function $h(t)$ which is also shown in the figure.}\label{fig e-folds}
\end{figure}

In the Fig. \ref{fig e-folds}.a, the evolution of $\epsilon$ and $\eta$ are shown. It has been also included the evolution of the Hubble parameter along with the torsion function and the field $\phi$ itself. Notice that assuming the above specified (well motivated) initial conditions, at the end of inflation ($\epsilon = 1$) we have the total e-folds $N \simeq 71$. But, from the uncertainty relation, the initial value of the field can be higher yielding a smaller number of e-folds during inflation. In this case, the field spends less time in the slow-roll regime which leads, therefore, to a decrease of $N$.

Although we have imposed $|\eta| \ll 1$ initially, this parameter grows very rapidly in the beginning of inflation and assumes a constant value ($\eta \simeq 0.3$) during almost the entire inflationary period. In the remaining 10 e-folds until the end of inflation $|\eta|$ is nearly zero, and finally it  becomes large in the last e-fold of inflation, corresponding to a substantial variation of $\epsilon$, which indicates the end of inflation. 

Notice that the change in the parameter $\eta$ is associated with the increase of the MDO field in the last 15 e-folds and also with a ``bump'' in the Hubble parameter in the same period. Such a bump, corresponding to a moderated increase in the energy density of the MDO field with respect to its initial value, is notably correlated with an increase of the torsion function $h(t)$. From the Eq. (\ref{torsion}) we see that initially, $|h(t_i)|$ is much smaller than the Hubble parameter, but it becomes of the same order of $H$ as $\phi \rightarrow \phi_c$. In the Fig. \ref{fig e-folds}.b it is also possible to note that the height of the bump and the corresponding transient increasing of $|h(t)|$ are strongly dependent on the value of $\phi_c$. The bump can even disappears for a value of $\phi_c$ small enough, but in this case changes in the initial conditions would be required in order to achieve the desired total e-folds of inflation. {An interesting question about the bump at the end of inflation is whether it could leave an imprint in the cosmic microwave background (CMB) radiation, since it is characterized by a sudden increase in the energy density at that time. A better quantitative treatment must be done in order to study the constraints in CMB spectrum due to such bump.}  

\begin{figure} 
\centering
\includegraphics[width=0.49\textwidth]{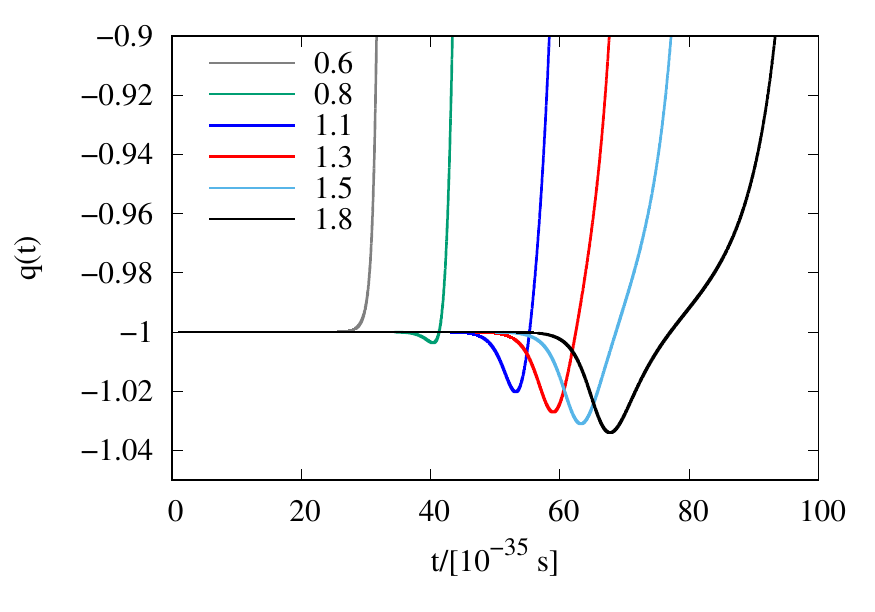}
\includegraphics[width=0.49\textwidth]{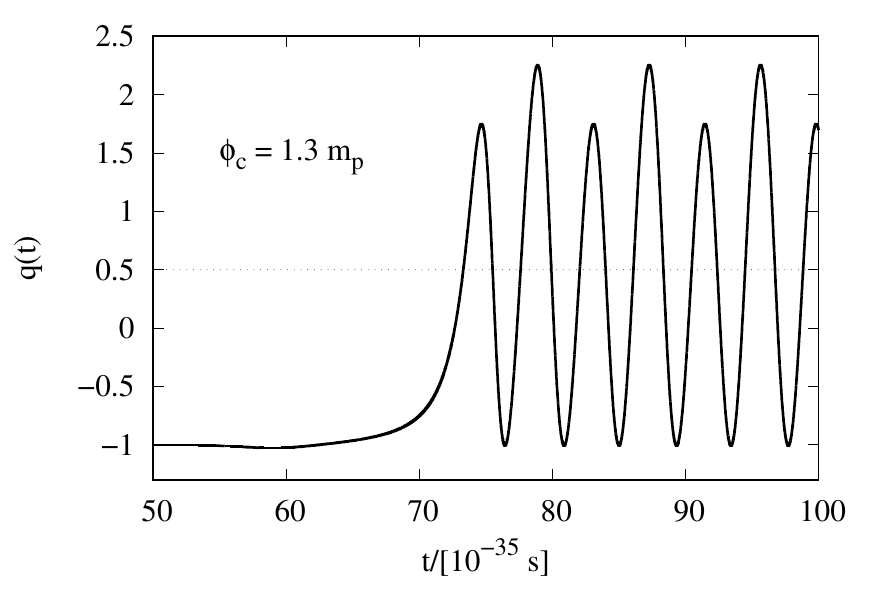}
\caption{{ Left panel: Evolution of the deceleration parameter $q(t)$ as a function of time at the final stage of inflation in the vicinity of the bump in the Hubble function (in the legend we show the values of the parameter $\phi_c$ in units of Planck mass). Notice that before the maximum of the bump we have $q(t) < -1$, i.e., a phantom like behavior. Right panel: Here we show the deceleration parameter for the first oscillations of the MDO field (for clarity these oscillations are not shown in the left panel). The field remains oscillating during all the reheating stage. The horizontal dotted line is the one period average of the deceleration parameter $\langle q(t) \rangle = 1/2$ indicating that the MDO field has a matter behavior in the reheating epoch as it happens for a typical massive scalar field for instance.}}\label{fig dec par}
\end{figure}

{ We have also studied the evolution of the deceleration parameter
\begin{equation}
q(t) = -\frac{\ddot{a}a}{\dot{a}^2} = -1-\frac{\dot{H}}{H^2},    
\end{equation}
at the inflationary period. In the left panel of the Fig. \ref{fig dec par} one can notice that $q(t) < -1$ for a short period preceding the maximum of the bump, for which we have again $q_{\rm max} = -1$ since $(\dot{H})_{\rm max} = 0$. Therefore, the present inflationary scenario for the MDO field exhibits a phase with a phantom like behavior. Again, it is explained by the increase of the function $|h(t)|$ which contributes significantly for the increase of the effective energy density at that epoch.}

After the end of inflation, the torsion remains important until the present time  and its contribution scales with the Hubble function $H(t)$, since the field acquires a constant value of the order of the Planck mass. Such a behavior is expected, since we are assuming the whole Universe homogeneously filled with the MDO field, and, therefore, each point of space containing a fermionic field interacts with torsion, contributing to the energy-momentum tensor and providing a contribution to the evolution. On the other hand, in the case of the chaotic inflation with a potential described by a sum of a quadratic and a quartic self-interacting term, the field goes to zero after inflation, thus the torsion which was initially important vanishes after inflation \cite{sra}. In this case, there is no bump at the final stage of inflation.

\section{Reheating}\label{sec 5}

{ After the end of inflation ($\epsilon = 1$ and $q = 0$) the amplitude of the MDO fermionic field $\phi(t)$ starts to oscillate coherently around the minimum of the potential with a frequency $\omega_\phi = \sqrt{V^{\prime\prime}(\phi)}$ as can be seen in the Fig. \ref{fig elko oscil} and a reheating mechanism takes place. During this period, the deceleration parameter oscillates around its average value $\langle q(t) \rangle = 1/2$ as shown in the right panel of the Fig. \ref{fig dec par}. Therefore, during the regime of coherent oscillations, the field behaves in average as non-relativistic particles, i.e. $\langle p_{\phi} \rangle = 0$, in the same way it happens for a typical scalar field subject to a quadratic potential. Such a result has already been verified in the Ref \cite{sra} for the MDO inflation with a quadratic mass term plus a self-interacting potential, and also for a symmetry breaking potential as can be seen in the Ref. \cite{st}. Now, let us briefly discuss two possible mechanisms for a successful reheating process after the MDO inflation, namely: \textbf{(i)} the electromagnetic coupling of the MDO field at high energies, and \textbf{(ii)} the coupling of the MDO field with the Higgs boson in a preheating phase preceding the reheating.}



Now, if the reheating process occurs with a MDO decay rate $\Gamma$  we need to add a damping term $\Gamma \dot{\phi}$ in the left-hand-side of the Eq. (\ref{phiElko}) which is negligible in the inflationary epoch. Therefore, the energy density of the MDO field is not conserved, but there is a term $-(1+\kappa^2\phi^2/8)\Gamma \dot{\phi}^2$ in the right-hand-side of the Eq. (\ref{phiElko2}). From the Eqs. (\ref{Elko density new}) and (\ref{Elko pressure new}) one can show that averaging in one period of oscillation $\langle (1+\kappa^2\phi^2/8) \dot{\phi}^2 \rangle = \rho_\phi$. Hence, by considering the mechanism \textbf{(i)}, the field decays in photons and the evolution of the MDO field and the decay product's energy density are described by the coupled equations
\begin{equation}
\dot{\rho}_\phi + 3H\rho_\phi=-\Gamma \rho_\phi,
\end{equation}
\begin{equation}
\dot{\rho}_r + 4H\rho_r=\Gamma\rho_\phi,
\end{equation}
where $\rho_r$ is the energy density of radiation. Notice that the above equations are identical in form to the case of reheating with a scalar field. { As a result while the MDO field dominates the energy density, the deceleration parameter is $1/2$ in average and the Universe is decelerating during all the reheating stage. Then, there is a transition to a radiation dominated epoch until the deceleration parameter reaches $q_{\rm rad} = 1$, which indicates the end of the reheating stage.}

In the Ref. \cite{alves2}, the interaction of the MDO field with photons at the tree level was studied. It was shown that electromagnetic interactions are excluded at the 95 \% CL by the LHC data for MDO masses up to 1 TeV. However, such a coupling might be relevant at the high energy levels achieved at the reheating phase {if we use the superior limit of mass discussed earlier, ${m}\simeq 2.4 \times 10^{10}$ GeV}. Let us suppose that the MDO decay rate is approximately given by $\Gamma \simeq g_e {m}$, where $g_e$ is a dimensionless coupling constant. In order to the damping term does not modify the evolution of the Universe in the inflationary epoch one has $\Gamma \ll H_i$, and than $g_e \ll 1.8$. If one considers the LHC upper bound $g_e \lesssim 10^{-5}$, for example, we have $\Gamma \lesssim 2.2 \times 10^5$ GeV. Therefore, in this reheating scenario, the Universe entropy can be generated by transferring the energy of the MDO field to photons. Then, the high energetic photons can generate other particles via pair production process.

The second possible mechanism for energy transfer after inflation is justified by the coupling of the MDO field with the Higgs field through the addition of the interaction term
\begin{equation}
{\mathcal L}_{\rm int} = h_e\Phi^\dag(x)\Phi(x)\stackrel{\neg}{\Lambda}(x)\Lambda(x)   
\end{equation}
in the Lagrangian density of the action (\ref{actionE}). Such an interaction has been considered to investigate the possibility of discovering the MDO field in the LHC  \cite{dias_lee,alves1,AlvesDias2014}. { Now, the full Lagrangian is given by
\begin{equation}
\mathcal{L} = \mathcal{L}_{\Lambda} + \mathcal{L}_{\rm int} + \mathcal{L}_{\Phi},
\end{equation}
where the last term is the Higgs Lagrangian. At the time of reheating, the MDO field oscillates as
\begin{equation}
\phi(t) = \phi_c + \phi_0(t)\cos(\omega_\phi t),    
\end{equation}
with a slowly varying amplitude $\phi_0(t)$ which is initially of the order of the Planck mass. Then, if we vary the above Lagrangian with respect to $\Phi$, one can obtain the equations of motion of the Higgs field coupled with $\phi(t)$, which will be similar to a Mathieu equation. The solutions of this equation are characterized by a parametric resonance which results in a fast amplification of $\Phi$. After such a period (called preheating) we have the reheating since the coupling of the Higgs field $\Phi$ with the standard-model particles leads to thermalization. Obviously, the temperature at the end of reheating depends on the details of the model.
}

In practice, both of the above described processes can happen simultaneously, although one might expect that one of them is dominant. Moreover, the success of the reheating process in the MDO cosmological scenario depends essentially in how strong are the couplings with the electromagnetic field and with the boson field. In this respect, a further investigation is necessary in order to constrain the parameter space of the models in a cosmological perspective.


\section{Conclusions}\label{sec 6}

In the present article we have studied an unified cosmological evolution driven by a MDO fermionic field. 

We have shown that the MDO inflation with a symmetry breaking potential can successfully achieve a number of e-folds as large as 71, with an initial condition for the field given by the smallest possible quantum fluctuation. The slow-roll regime is maintained during the whole inflation with a constant parameter $\eta \simeq 0.3$ during almost the entire period. The torsion function plays and important role in the remaining 15 e-folds until the end of inflation generating a bump in the Hubble function at that epoch.

Such a sudden increase of the energy density of the Universe to a value greater than the initial value rises the question if it could generate any imprint in the cosmic microwave background anisotropies spectra. Similarly, one may expect a modification  in the form of the primordial spectrum of gravitational waves. These issues will be answered by forthcoming investigations.

It is worth to stress that the dark matter behavior of the MDO fermionic field in the way it is understood in the present article is quite different from the preceding works on the topic \cite{st,sra}. In those works, the coherent oscillations of the field after inflation is interpreted as dark matter. This is because the oscillating field is pressureless in average in the same manner it happens with an oscillating scalar field subject to a quadratic potential. However, this is possible only in the case the MDO field interaction with any other field is negligibly small. Conversely, if the field couples with radiation or with the Higgs boson at the still high energetic regime after the inflationary epoch, than its energy density will decay exponentially in a reheating phase as discussed in general in the Section \ref{sec 5}. In this sense, the MDO dark matter behavior only survives if the field acquires a non-null constant value after reheating which is of the order of the Planck mass or, more precisely, $\phi_c = m_{pl} \sqrt{\Omega_{{\rm DM},\phi}/\pi\Omega_b}$. Therefore, if this condition is satisfied, it does not matter what kind of potential the field is subject (neither the inflationary mechanism), the gravitational coupling of the field with baryonic matter leads to a contribution in the form of a pressureless matter. It is worth to emphasize that such a coupling has its origin in the definition of the spin connections in curved spacetimes. No explicit coupling between the MDO field with other fields were assumed.

Moreover, if the potential is not null in $\phi=\phi_c$, than the MDO field has an additional constant contribution to the energy density that works exactly as a cosmological constant. {Such term is also responsible to constrain the mass of the field in the range $0 < m \lesssim 2.4 \times 10^{10}$ GeV.}

Finally, the scenario described in this article provide us with a natural explanation for the cosmological coincidence problem. This is because the evolution of the energy density of the MDO field scales with the energy density evolution of other matter fields, resulting in a present value of the same order of magnitude of the usual matter. 

Therefore, we conclude that MDO fermionic field is a good candidate to drive the whole evolution of the Universe in an unified fashion, in such a way that the inflationary field, dark matter and dark energy are described by different manifestations of a single fermionic field.  Additionally, as already pointed out at the Introduction, a possible interpretation of the inflationary phase as a consequence of the Pauli exclusion principle put this model on very interesting physical grounds.

\begin{acknowledgements}
This study was financed in part by the Coordena\c{c}\~ao de Aperfei\c{c}oamento de Pessoal de N\'ivel Superior - Brasil (CAPES) - Finance Code 001. SHP would like to thank CNPq - Conselho Nacional de Desenvolvimento Cient\'ifico e Tecnol\'ogico, Brazilian research agency, for financial support, grants number 303583/2018-5 and 400924/2016-1. MESA would like to thank the Brazilian agency FAPESP for financial support under the thematic project \# 13/26258-4. TMG would like to thank CAPES for financial suport.
\end{acknowledgements}


\begin{thebibliography}{99}
\bibitem{AHL1} D. V. Ahluwalia-Khalilova and D. Grumiller, 
{Dark matter: A spin one-half fermion field with mass dimension one?}, 
\emph{Phys. Rev. D} {\bf 72} (2005) 067701. 

\bibitem{AHL2} D. V. Ahluwalia-Khalilova and D. Grumiller, 
{Spin half fermions with mass dimension one: theory, phenomenology, and dark matter}, 
\textit{JCAP} \textbf{07} (2005) 012.

\bibitem{ahl2011a}D. V. Ahluwalia, C.-Y. Lee, and D. Schritt, 
{Self-interacting Elko dark matter with an axis of locality}, 
\emph{Phys. Rev. D} {\bf83} (2011) 065017.


\bibitem{AHL4} D. V. Ahluwalia, 
{The Theory of Local Mass Dimension One Fermions of Spin One Half}, 
\textit{Adv. Appl. Clifford Algebras} (2017) 1-39.


\bibitem{AHL3} D. V. Ahluwalia, 
{Evading Weinberg's no-go theorem to construct mass dimension one fermions: Constructing darkness},
\textit{Europhys. Lett.} {\bf118} (2017) 60001.


\bibitem{alves1} A. Alves, F. de Campos, M. Dias and J. M. Hoff da Silva, 
{Searching for Elko dark matter spinors at the CERN LHC},  
\emph{Int. J. Mod. Phys.} {\bf A 30} (2015) 1550006.

\bibitem{dias_lee} M. Dias and Cheng-Yang Lee,
{Constraints on mass dimension one fermionic dark matter from the Yukawa interaction},
\emph{Phys. Rev. D} {\bf94} (2016) 065020. 

\bibitem{AlvesDias2014} A. Alves, M. Dias and F. de Campos, 
{Perspectives for an Elko phenomenology using monojets at the 14 TeV LHC},
\emph{Int. J. Mod. Physics D} \textbf{14} (2014)1 444005.

\bibitem{alves2} A. Alves, M. Dias, F. de Campos, L. Duarte and J. M. Hoff da Silva,
{Constraining Elko dark matter at the LHC with monophoton events},
\emph{Europhys. Lett.} \textbf{121} (2018) 31001.

\bibitem{FABBRI} L. Fabbri, 
{The most general cosmological dynamics for Elko matter fields}, 
\emph{Phys. Lett. B} {\bf704} (2011) 255.

\bibitem{BOE4} C. G. Boehmer, 
{Dark spinor inflation - theory primer and dynamics}, 
\emph{Phys. Rev. D} {\bf77} (2008) 123535.

\bibitem{BOE6} C. G. Boehmer, J. Burnett, D. F. Mota and D. J. Shaw, 
{Dark spinor models in gravitation and cosmology}, 
\textit{JHEP} \textbf{07} (2010) 053.

\bibitem{GREDAT} D. Gredat and S. Shankaranarayanan, 
{Modified scalar and tensor spectra in spinor driven inflation}, 
\textit{JCAP} \textbf{01} (2010) 008.

\bibitem{BASAK} A. Basak and J. R. Bhatt, 
{Lorentz invariant dark-spinor and inflation}, 
\textit{JCAP} \textbf{06} (2011) 011.

\bibitem{basak2015} {A. Basak and S. Shankaranarayanan, 
{Super-inflation and generation of first order vector perturbations in ELKO},
\textit{JCAP} \textbf{05} (2015) 034.}

\bibitem{sadja} H. M. Sadjadi, 
{On coincidence problem in Elko dark energy model}, 
\emph{Gen. Relativ. Gravit.} {\bf 44} (2012) 2329.

\bibitem{saj} S. H. Pereira, A. Pinho S. S., J. M. Hoff da Silva, 
{Some remarks on the attractor behaviour in Elko cosmology}, 
\textit{JCAP} \textbf{08} (2014) 020.

\bibitem{js} J. M. Hoff da Silva and S. H. Pereira, 
{Exact solutions to Elko spinors in spatially flat Friedmann-Robertson-Walker spacetimes}, 
\textit{JCAP} {\bf 03} (2014) 009.

\bibitem{asf} A. Pinho S. S., S. H. Pereira, J. F. Jesus, 
{A new approach on the stability analysis in Elko cosmology}, 
\textit{Eur. Phys. J. C.} {\bf 75} (2015) 36.

\bibitem{kouwn} S. Kouwn, J. Lee, T. H. Lee and P. Oh, 
{Elko spinor model  with torsion and cosmology}, 
\emph{Mod. Phys. Lett. A} {\bf28} (2013) 1350121.

\bibitem{sajf} S. H. Pereira, A. Pinho S. S., J. M. Hoff da Silva and J. F. Jesus, 
{$\Lambda(t)$ cosmology induced by a slowly varying Elko field}, 
\textit{JCAP} {\bf 01} (2017) 055.

\bibitem{st} S. H. Pereira and T. M. Guimar\~aes, 
{From inflation to recent cosmic acceleration: the fermionic Elko field driving the evolution of the universe},
\textit{JCAP} {\bf 09} (2017) 038.

\bibitem{sra} S. H. Pereira, R. F. L. Holanda and A. Pinho S. S.,
{Evolution of the universe driven by a mass-dimension-one fermion field},
\text{Europhys. Lett.} {\bf 120} (2017) 31001.

\bibitem{dias1} M. Dias, F. de Campos and J. M. Hoff da Silva, 
{Exploring Elko typical signature}, 
\emph{Phys. Lett. B} {\bf706} (2012) 352. 

\bibitem{elkosigma} R. J. Bueno Rog\' erio, J. M. Hoff da Silva, S. H. Pereira and Rold\~ao da Rocha, 
{A framework to a mass-dimension-one fermionic sigma model},
\emph{Europhys. Lett.} \textbf{113} (2016) 60001.

\bibitem{elkocasimir}S. H. Pereira, J. M. Hoff da Silva and R. dos Santos,
{Casimir effect for Elko fields},
\emph{Mod. Phys. Lett.} {\bf A 32} (2017) 1730016.

\bibitem{elkoeffective}R. J. Bueno Rogerio, J.M. Hoff da Silva, M. Dias and S.H. Pereira, 
{Effective lagrangian for a mass dimension one fermionic field in curved spacetime},
\textit{JHEP}, {\bf 02} (2018) 145.

\bibitem{sarich} S. H. Pereira and R. S. Costa,
{Partition function for a mass dimension one fermionic field and the dark matter halo of galaxies},
\emph{arXiv:1807.06944 [physics.gen-ph]}.

\bibitem{Liddle2006} A.R. Liddle and L.A. Ure\~na-L\'opez, 
{Inflation, Dark Matter, and Dark Energy in the String Landscape},
\emph{Phys. Rev. Letters} \textbf{97} (2006) 161301.

\bibitem{Liddle2008} A.R. Liddle, C. Pahud and L.A. Ure\~na-L\'opez,
{Triple unification of inflation, dark matter, and dark energy using a single field},
\emph{Phys. Rev. D} \textbf{77} (2008) 121301.

\bibitem{Bastero2016} M. Bastero-Gil, R. Cerezo and J.G. Rosa,
{Inflaton dark matter from incomplete decay},
\emph{Phys. Rev. D} \textbf{93} (2016) 103531.

\bibitem{bookliddle}A. R. Liddle and D. H. Lyth, 
{Cosmological inflation and large-scale structure}, 
Cambridge University Press, U.K, (2000) [ISBN:9781108004916].

\bibitem{linde1} A. D. Linde, 
{Inflationary Cosmology after Planck 2013}, 
(2014), arXiv:1402.0526 [hep-th].
 
\bibitem{kolb} E. W. Kolb and M. S. Turner, {\it The Early Universe}, Westview Press, U.S.A, (1990) [ISBN:0201626748].

\bibitem{saha1997} B. Saha and G.N. Shikin, 
{Interacting Spinor and Scalar Fields in Bianchi Type I Universe Filled with Perfect Fluid: Exact Self-Consistent Solutions},
\emph{Gen. Relativ. Gravit.} \textbf{29} (1997) 1099.

\bibitem{saha2004} B. Saha and T. Boyadjiev, 
{Bianchi type-I cosmology with scalar and spinor fields},
\emph{Phys. Rev. D} {\bf 69} (2004)  124010.

\bibitem{saha2006} B. Saha, 
{Nonlinear spinor field in Bianchi type-I cosmology: Inflation, isotropization, and late time acceleration},
\emph{Phys. Rev. D} {\bf 74} (2006) 124030.

\bibitem{grams2014} G. Grams, R.C. de Souza and G.M. Kremer,
{Fermion field as inflaton, dark energy and dark matter}, 
\emph{Class. Quantum Grav.} {\bf 31} (2014) 185008. 
 
\bibitem{tsam} M. Tsamparlis, 
{Cosmological principle and torsion}, 
\textit{Phys. Lett. A} {\bf 75} (1979) 27.

\bibitem{Aghanim2018} Planck Collaboration: N. Aghanim et al. 
{Planck 2018 results. VI. Cosmological parameters}, 
(2018) arXiv: 1807.06209 .

\bibitem{Linde1993} A. Linde, {Hybrid inflation}, \textit{Phys. Rev. D} {\bf 49} (1994), 748.



\end{thebibliography}

\end{document}